\documentclass[appendixfloats]{emulateapj}
\usepackage{natbib}
\usepackage{apjfonts}
\usepackage{graphicx}

\begin{document}

\shorttitle{Massive galaxy clustering}
\shortauthors{White et al.}

\title{The clustering of massive galaxies at $\lowercase{z}\sim 0.5$ from the
first semester of BOSS data}
\author{Martin White${}^{1,2,3}$,
M. Blanton${}^4$, A. Bolton${}^5$, D. Schlegel${}^3$, J. Tinker${}^4$,
A. Berlind${}^6$, L. da Costa${}^7$, E. Kazin${}^4$, Y.-T. Lin${}^8$,
M. Maia${}^7$, C.K. McBride${}^6$, N. Padmanabhan${}^9$,
J. Parejko${}^9$, W. Percival${}^{10}$, F. Prada${}^{11}$, B. Ramos${}^7$,
E. Sheldon${}^{12}$, F. de Simoni${}^7$, R. Skibba${}^{13}$,
D. Thomas${}^{10}$, D. Wake${}^9$, I. Zehavi${}^{14}$, Z. Zheng${}^9$,
R. Nichol${}^{10}$, Donald P. Schneider${}^{15}$,
Michael A. Strauss${}^{16}$, B.A. Weaver${}^4$,
David H. Weinberg${}^{17}$
}
\affil{${}^1$ Department of Physics,
University of California Berkeley, CA}
\affil{${}^2$ Department of Astronomy,
University of California Berkeley, CA}
\affil{${}^3$ Lawrence Berkeley National Laboratory,
1 Cyclotron Road, Berkeley, CA}
\affil{${}^4$ Center for Cosmology and Particle Physics,
New York University, NY}
\affil{${}^5$ Dept. Physics and Astronomy, University of Utah, UT}
\affil{${}^6$ Dept. Physics, Vanderbilt University, Nashville, TN}
\affil{${}^7$ Observatorio Nacional, Brazil}
\affil{${}^8$ IPMU, University of Tokyo, Japan}
\affil{${}^9$ Yale Center for Astronomy and Astrophysics, Yale University,
New Haven, CT}
\affil{${}^{10}$ Institute of Cosmology and Gravitation, University of
Portsmouth, UK}
\affil{${}^{11}$ Instituto de Astrofisica de Andalucia, Granada, Spain}
\affil{${}^{12}$ Brookhaven National Laboratory}
\affil{${}^{13}$ Steward Observatory, University of Arizona, AZ}
\affil{${}^{14}$ Dept. Astronomy, Case Western Reserve University, OH}
\affil{${}^{15}$ Dept. Astronomy and Astrophysics, Penn State University, PA}
\affil{${}^{16}$ Dept. Astrophysical Sciences, Princeton University, NJ}
\affil{${}^{17}$ Ohio State University, Dept. of Astronomy and CCAPP,
Columbus, OH}

\date{\today}

\begin{abstract}
We calculate the real- and redshift-space clustering of massive galaxies
at $z\sim 0.5$ using the first semester of data by the Baryon Oscillation
Spectroscopic Survey (BOSS).
We study the correlation functions of a sample of 44,000 massive galaxies
in the redshift range $0.4<z<0.7$.
We present a halo-occupation distribution modeling of the clustering results
and discuss the implications for the manner in which massive galaxies at
$z\sim 0.5$ occupy dark matter halos.
The majority of our galaxies are central galaxies living in halos of mass
$10^{13}\,h^{-1}M_\odot$, but $10\%$ are satellites living in halos 10 times
more massive.
These results are broadly in agreement with earlier investigations of massive
galaxies at $z\sim 0.5$.  The inferred large-scale bias ($b\simeq 2$) and
relatively high number density
($\bar{n}=3\times 10^{-4}\,h^3\,{\rm Mpc}^{-3}$)
imply that BOSS galaxies are excellent tracers of large-scale structure,
suggesting BOSS will enable a wide range of investigations on the distance
scale, the growth of large-scale structure, massive galaxy evolution and
other topics.
\end{abstract}


\section{Introduction}

The distribution of objects in the Universe displays a high degree of
organization, which in current models is due to primordial fluctuations
in density which were laid down at very early times and amplified by the
process of gravitational instability.
Characterizing the evolution of this large-scale structure is a central
theme of cosmology and astrophysics.
In addition to allowing us to understand the structure itself, large-scale
structure studies offer an incisive tool for probing cosmology and particle
physics and sets the context for our modern understanding of galaxy formation
and evolution.
Since the pioneering studies of \citet{Hum56,GreTho78,JoeEin78} and the first
CfA redshift survey \citep{Huc83}, galaxy redshift surveys have played a key
role in this enterprise, and ever larger surveys have provided increasing
insight and ever tighter constraints on cosmological models.

This paper presents the first measurements of the clustering of massive
galaxies from the Baryon Oscillation Spectroscopic Survey
\citep[BOSS;][]{SchWhiEis09}
based on a sample of galaxy redshifts observed during the period January
through July 2010.
We demonstrate that BOSS is efficiently obtaining redshifts of some of the
most luminous galaxies at $z\simeq 0.5$, and has already become the largest
such redshift survey ever undertaken.  The high bias and number density of
these objects (described below) make them ideal tracers of large-scale
structure, and suggest that BOSS will make a significant impact on many
science questions including a determination of the cosmic distance scale,
the growth of structure and the evolution of massive galaxies.

The outline of the paper is as follows.
In \S\ref{sec:obs} we briefly describe the BOSS survey and observations,
and define the sample we focus on in this paper.  Our clustering results
are described in \S\ref{sec:clustering} and interpreted in the framework
of the halo model in \S\ref{sec:hod}, where we also compare to previous
work on the clustering of massive galaxies at intermediate redshift.
We conclude with a discussion of the implications of these results in
\S\ref{sec:discussion}, while some technical details on the construction
of our mock catalogs are relegated to an Appendix.
Throughout this paper when measuring distances we refer to comoving
separations, measured in $h^{-1}$Mpc with
$H_0=100\,h\,{\rm km}\,{\rm s}^{-1}\,{\rm Mpc}^{-1}$.
We convert redshifts to distances, assuming a $\Lambda$CDM cosmology
with $\Omega_m=0.274$, $\Omega_\Lambda=0.726$ and $h=0.70$.  This is
the same cosmology as assumed for the N-body simulations from which we
make our mock catalogs (see Appendix \ref{app:mocks}).

\section{Observations} \label{sec:obs}

The Sloan Digital Sky Survey \citep[SDSS;][]{Yor00} mapped nearly a quarter
of the sky using the dedicated Sloan Foundation 2.5 m telescope \citep{Gun06}
located at Apache Point Observatory in New Mexico.
A drift-scanning mosaic CCD camera \citep{Gun98} imaged the sky in
five photometric bandpasses \citep{Fuk96,Smi02,Doi10}
to a limiting magnitude of $r\simeq 22.5$.
The imaging data were processed through a series of pipelines that perform
astrometric calibration \citep{Pie03},
photometric reduction \citep{Photo},
and photometric calibration \citep{Pad08}.
The magnitudes were corrected for Galactic extinction using the maps of
\citet{SFD98}.
BOSS, a part of the SDSS-III survey (Eisenstein et al., in prep.) has
completed an additional $3,000$ square degrees of imaging in the southern
Galactic cap, taken in a manner identical to the original SDSS imaging.
All of the data have been processed through the latest versions of the
pipelines and BOSS is obtaining spectra of a selected subset
(Padmanabhan et al., in preparation) of 1.5 million galaxies approximately
volume limited to $z\simeq 0.6$
(in addition to spectra of 150,000 quasars and various ancillary observations).
The targets are assigned to tiles of diameter $3^\circ$ using an
adaptive tiling algorithm \citep{Tiling}.  Aluminum plates are drilled
with holes corresponding to the positions of objects on each tile, and
manually plugged with optical fibers that feed a pair of double
spectrographs.  These spectrographs are significantly upgraded from
those used by SDSS-I/II \citep{Yor00,Sto02}, with improved chips with
better red response, higher throughput gratings, 1,000 fibers (instead of
640) and a $2''$ entrance aperture (was $3''$).
The spectra cover the range $3,600\,$\AA\ to $10,000\,$\AA, at a resolution of
about 2,000.

\begin{figure}
\begin{center}
\resizebox{3.2in}{!}{\includegraphics{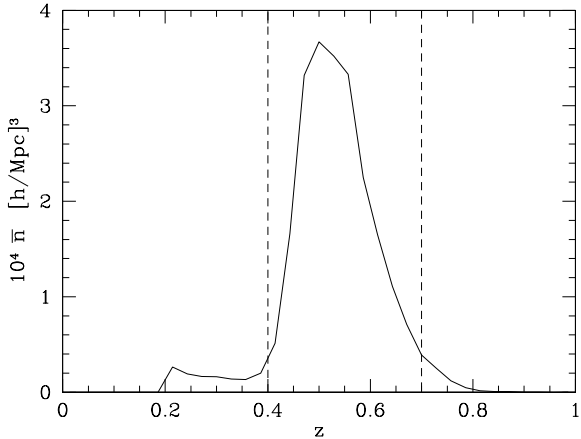}}
\end{center}
\caption{The (comoving) number density of galaxies, $\bar{n}(z)$ for the
sample described in the text (\S\protect\ref{sec:obs}).  The vertical dashed
lines indicate the redshift limits we use in our analysis: $0.4<z<0.7$.}
\label{fig:nbar}
\end{figure}

\begin{figure}
\begin{center}
\resizebox{3.2in}{!}{\includegraphics{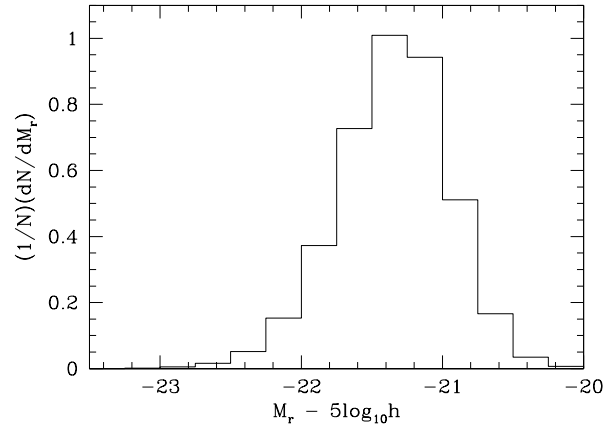}}
\end{center}
\caption{The distribution of absolute magnitudes for the sample
analyzed in this paper.  We have $k+e$ corrected the $r$-band magnitudes to
$z\simeq 0.55$ using the $g-i$ color assuming a passively evolving
galaxy -- since the redshift range is small this amounts to a small
correction.  This sample consists of intrinsically very bright, and massive,
galaxies with stellar masses several times the characteristic mass in a
Schechter fit.  The luminosity function of \protect\citet{Fab07} at $z=0.5$,
converted from $B$ to $r$ band assuming a redshifted $z=0$ elliptical galaxy
template, has a characteristic luminosity of $-19.8$.
Converting the \protect\citet{Bel04} luminosity function using a high-$z$
single burst model gives $-20$.  So all of the CMASS galaxies are brighter
than this characteristic luminosity.}
\label{fig:maghist}
\end{figure}

BOSS makes use of luminous galaxies selected from the multi-color SDSS
imaging to probe large-scale structure at intermediate redshift ($z<0.7$).
These galaxies are among the most luminous galaxies in the universe and trace
a large cosmological volume while having high enough number density to ensure
that shot-noise is not a dominant contributor to the clustering variance.
The majority of the galaxies have old stellar systems whose prominent
$4,000\,$\AA\ break in their spectral energy distributions makes them relatively
easy to select in multi-color data.

The strategy behind, and details of, our target selection are covered in
detail in Padmanabhan et al.~(in preparation).
Cuts in color-magnitude space allow a roughly volume-limited sample of
luminous galaxies to be selected, and partitioned into broad redshift bins.
Briefly, we follow the SDSS-I/II procedure described in \citet{Eis01} and
define a ``rotated'' combination of colors
%\begin{eqnarray}
%  c_\perp &=& (r-i)-(g-r)/4-0.18 \\
%  c_{||}  &=& 0.7(g-r)+1.2(r-i-0.18) \\
%  d_\perp &=& (r-i) - (g-r)/8
%\end{eqnarray}
$d_\perp=(r-i)-(g-r)/8$.
The sample we analyze in this paper (the so-called ``CMASS sample'' since it
is approximately stellar mass limited) is defined via
\begin{equation}
\begin{array}{c}
  d_\perp>0.55 \quad{\rm and}\quad
  i < 19.9 \quad {\rm and}\quad i_{\rm fiber2} < 21.5 \\
  i < 19.86+1.6\,(d_\perp-0.8) \quad {\rm and} \quad r-i<2
\end{array}
\label{eqn:cuts}
\end{equation}
where magnitude cuts use ``cmodel magnitudes'' and colors are defined with
``model magnitudes'', except for $i_{\rm fiber2}$ which is the magnitude
in the $2''$ spectroscopic fiber
\citep[see][for definitions of the magnitudes and further discussion]{Sto02,DR2}.
There are two additional cuts to reduce stellar contamination,
$z_{\rm psf}-z\ge 9.125-0.45\,z$ and
$r_{\rm psf}-r>0.3$.

These cuts isolate the $z\sim 0.5$, high mass galaxies.
The $i-d_\perp$ constraint is approximately a cut in absolute magnitude or
stellar mass, with $d_\perp$ closely tracking redshift for these galaxies.
As discussed in detail in Padmanabhan et al.~(in preparation), the slope of the
$i-d_\perp$ cut is set to parallel the track of a passively evolving,
constant stellar mass galaxy as determined from the population synthesis
models of \citet{Mar09}.
This approach leads to an approximately stellar mass limited sample.
We restrict ourselves to galaxies in the redshift range $0.4<z<0.7$
(Fig.~\ref{fig:nbar}).  Note that our selection gives the majority of the
galaxies within $\Delta z=0.1$ of the median -- this has the advantage of
making the analysis relatively straightforward but means we need to combine
with other samples to obtain leverage in redshift.
A comparison of the cuts defining this sample with other, similar, samples in
the literature will be presented in Padmanabhan et al.~(in prep.).
In general BOSS goes both fainter and bluer than the earlier samples,
targeting ``luminous galaxies'' not ``luminous red galaxies''.

The distribution of absolute ($r$-band) magnitude for the sample is shown
in Fig.~\ref{fig:maghist}, where we see that all of the CMASS galaxies are
intrinsically very luminous.  Using the modeling of Maraston et al.~(in
preparation) on the BOSS spectra we find the median stellar mass of the
sample is $10^{11.7}\,M_\odot$.
While the detailed numbers depend on assumptions about e.g.~the initial
mass function, these galaxies are at the very high mass end of the stellar
mass function at this redshift for any reasonable assumptions.

\begin{figure}
\begin{center}
\resizebox{3.2in}{!}{\includegraphics{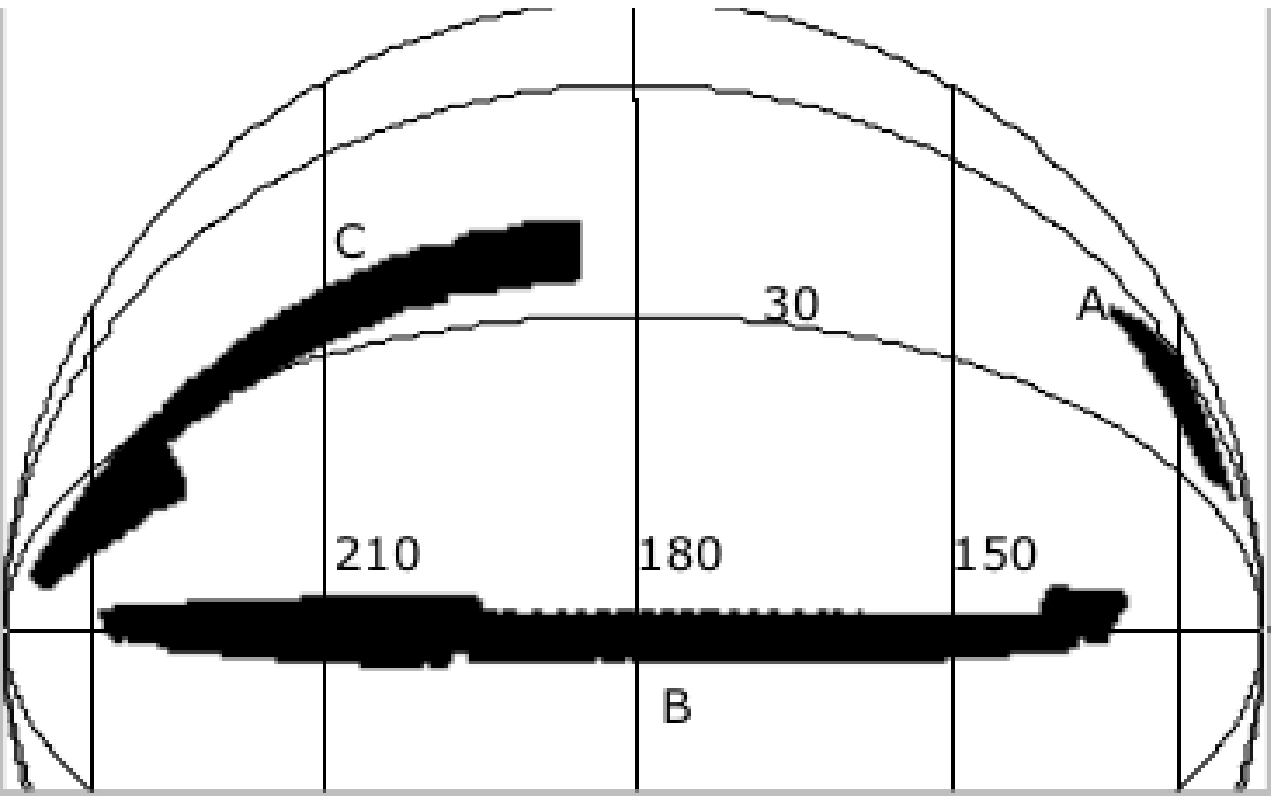}}
\resizebox{3.6in}{!}{\includegraphics{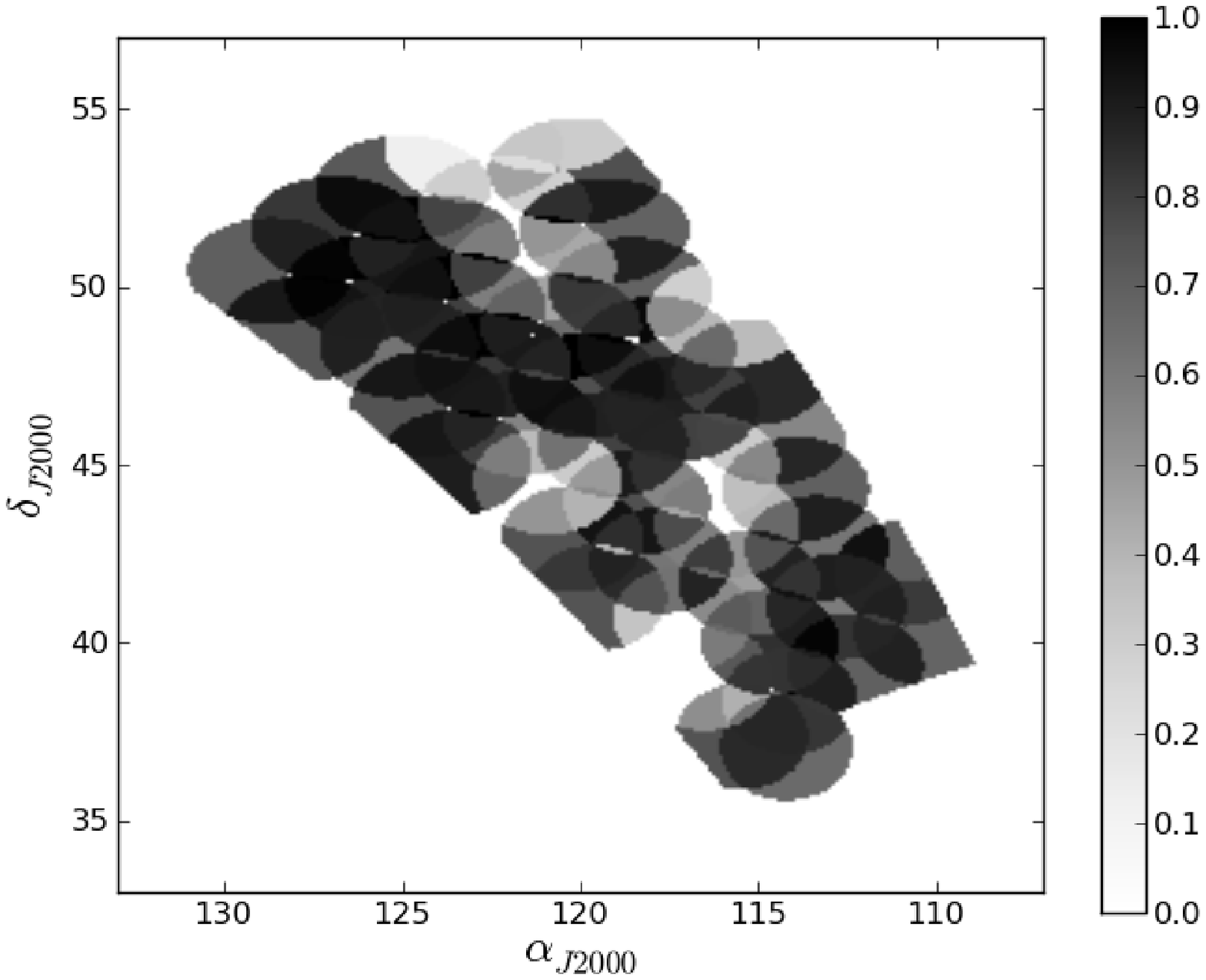}}
\end{center}
\caption{(Top) The sky coverage of the galaxies used in this analysis, in
orthographic projection centered on $\alpha_{J2000}=180^\circ$ and
$\delta_{J2000}=0^\circ$.
The regions A, B and C described in the text are marked.
(Bottom) A zoom in of region A with the greyscale showing completeness.
This region is the most contiguous of the three, and region B is the least
contiguous owing to hardware problems in the early part of the year.}
\label{fig:sky}
\end{figure}

%\begin{table}
%\begin{center}
%\begin{tabular}{cl}
%Spectro& $r<19.2$,\ $r<13.1+c_{||}/0.3$, \\
%Cut I  & $|c_\perp|<0.2$ \\ \hline
%Spectro& $r<19.5$,\quad $|c_\perp|>0.45-(g-r)/6$,\\
%Cut II & $g-r>1.30+0.25(r-i)$ \\ \hline
%Photo  & $r<19.7$,\ $r<13.6+c_{||}/0.3$ \\
%Cut I  & $|c_\perp|<0.2$, \\ \hline
%Photo  & $i<20$,\quad $|d_\perp|>0.55$,\\
%Cut II & $i<18.3+2d_\perp$,\quad $g-r>1.4$ \\ \hline
%2SLAQ  & $i<19.8$,\quad $|d_\perp|>0.65$,\\
%       & $c_{||}\ge 1.6$,\quad $r-i<2$
%\end{tabular}
%\end{center}
%\caption{{\rm A table comparing the selections of the different LRG samples.
%The lines labeled Spectro refer to the spectroscopic LRG sample of
%\protect\citet{Eis01}, those labeled Photo are for the photometric
%LRG sample described in \protect\citet{Pad05} and those labeled 2SLAQ
%are for the sample described in \protect\citet{Can06}.  These selections
%use in addition to $d_\perp$ defined in the text the rotated combinations
%of colors $c_{||} = 0.7\,(g-r)+1.2\,(r-i-0.18)$ and
%$c_\perp = (r-i)-(g-r)/4-0.18$.
%The corresponding cuts for our sample are given in
%Eq.~(\protect\ref{eqn:cuts}).}}
%\label{tab:samples}
%\end{table}

The clustering measurements in this paper are based on the data taken by
BOSS up to end of July 2010, which includes 120,000 galaxies over
1,600 deg${}^2$ of sky.  However, the data prior to January 2010 were taken in
commissioning mode and little of those data are of survey quality.  Once we
trim the data to contiguous regions (Fig.~\ref{fig:sky}) with high redshift
completeness and select galaxies at $z\sim 0.5$ we are left with $44,000$
galaxies, covering $580$ square degrees, which we have used in our analysis.

The sky coverage of our sample can be seen in Figure \ref{fig:sky}.
We view the data as comprising three regions of the sky, hereafter referred
to as A, B and C (see Figure).  Galaxies in these regions are separated from
those in any other region by several hundred Mpc, and we shall consider them
independent.  Convenient ``rectangular'' boundaries to the regions are
\begin{eqnarray}
  A &:& 105^\circ<\alpha_{J2000}<135^\circ \quad , \quad
         25^\circ<\delta_{J2000}< 60^\circ \\
  B &:& 125^\circ<\alpha_{J2000}<240^\circ \quad , \quad
         -5^\circ<\delta_{J2000}<5^\circ  \\
  C &:& 185^\circ<\alpha_{J2000}<255^\circ \quad , \quad
         10^\circ<\delta_{J2000}<45^\circ \quad .
\end{eqnarray}
These boundaries yield widths (heights) of 600 (700), 2600 (270) and 1600 (800)
$h^{-1}$Mpc respectively at $z\simeq 0.5$.
As we shall discuss below, the data are consistent with having the same
clustering and redshift distribution in all three regions.

\section{Clustering measures} \label{sec:clustering}

\begin{figure}
\begin{center}
\resizebox{3.2in}{3.1in}{\includegraphics{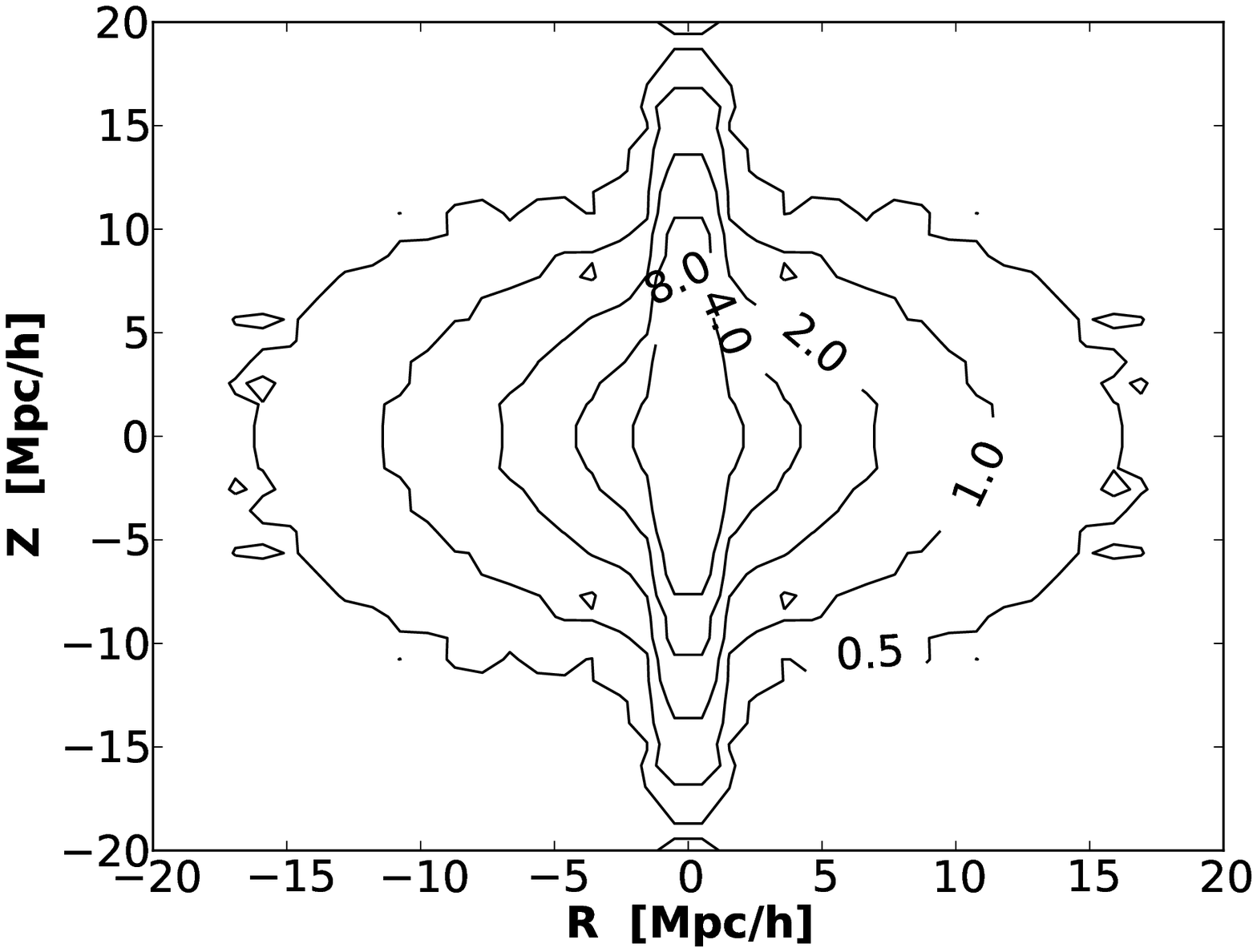}}
\resizebox{3.2in}{3.1in}{\includegraphics{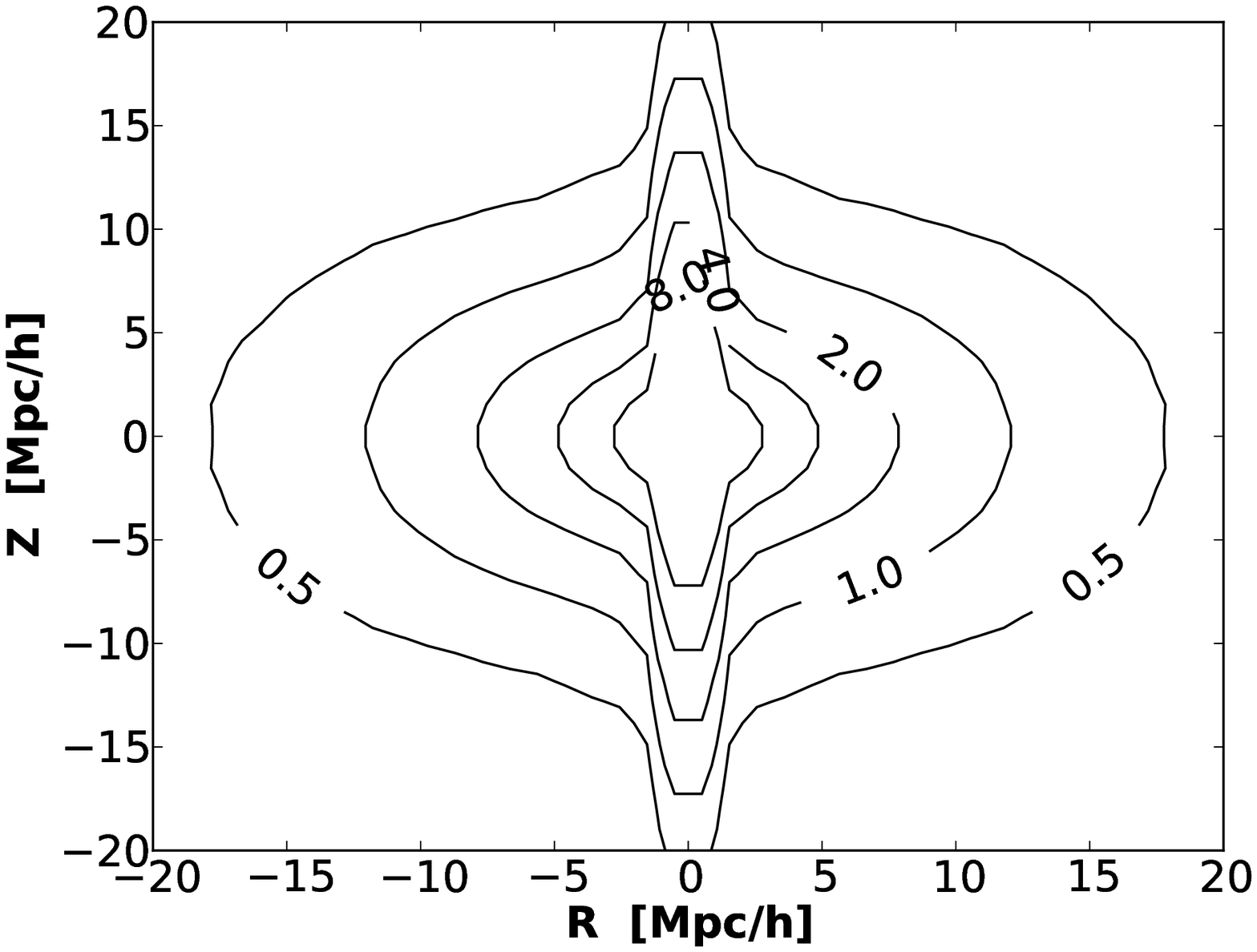}}
\end{center}
\caption{Contours of the redshift-space correlation function, $\xi(R,Z)$,
for our $0.4<z<0.7$ galaxy sample (see text).  Note the characteristic
elongation in the $Z$ direction at small $R$ (fingers-of-god) and squashing
at large $R$ (super-cluster infall).  The upper panel shows the results from
the BOSS data, while the lower panel is from our mock catalogs.  The level
of agreement is quite good, as can be seen more quantitatively in later
figures.}
\label{fig:xiRZ}
\end{figure}

We compute several two-point, configuration-space clustering statistics in
this paper.
The basis for all of these calculations is the two-point galaxy correlation
function on a two-dimensional grid of pair separations parallel and
perpendicular to the line-of-sight: $\xi(R,Z)$.

To estimate the counts expected for unclustered objects while accounting
for the complex survey geometry, we generate random catalogs with the
detailed radial and angular selection functions of the sample but
with $50\times$ the number of points.
Numerous tests have confirmed that the survey selection function factorizes
into an angular and a redshift piece.
The redshift selection function can be taken into account by distributing
the randoms according to the observed redshift distribution of the sample.
The completeness on the sky is determined from the fraction of target galaxies
in a sector for which we obtained a high-quality redshift, with the sectors
being areas of the sky covered by a unique set of spectroscopic tiles
\citep[see][]{Tiling,Teg04}.
We use the {\sc Mangle\/} software \citep{Mangle} to track the angular
completeness.
In computing the redshift completeness we omit galaxies for which a redshift
was already known from an earlier survey from both the target and success lists,
and then later randomly sample such galaxies with the resulting completeness in
constructing the input catalog.  Since very few of our targets at $z\sim 0.5$
have existing redshifts this is a very small correction.
Not all of the spectra taken resulted in a reliable redshift, and the failure
probability has angular structure due to hardware limitations.
These result in spatial signal-to-noise fluctuations in observations.
We find no evidence that this failure is redshift dependent - low and high
redshift failure regions have the same redshift distribution.
We therefore apply a small angular correction for this spatial structure by
up-weighting galaxies based on the signal-to-noise of each spectrum, and the
probability of redshift measurement. This is a small correction and only
affects our results at the percent level.
To avoid issues arising from small-number statistics we only keep sectors
with area larger than $10^{-4}\,$sr, or approximately $0.3$ sq.~deg.  At the
observed mean density ($150\,{\rm deg}^{-2}$) we expect several tens of
galaxies in any such region,
enabling us to reliably determine the redshift completeness\footnote{Assuming
binomial statistics, if $M$ of $N$ galaxies have redshifts the most likely
completeness is $c=M/N$, the mean is $(M+1)/(N+2)$ and the variance on $c$ is
$[M(N-M)+N+1]/[(N+2)^2(N+3)]$.  For example if $N=12$ and $M=9$ the error on
$c$ is approximately $10\%$.  For $N\ge 100$ the error is under $5\%$.
Unless the scatter is somehow correlated with the signal these uncertainties
are negligible.  In fact, we find that ignoring the exact value of the
completeness in constructing our random catalog only slightly alters our
final $\xi$.}.
We trim the final area to all sectors with completeness greater than 75\%,
producing our final sample of 44,000 galaxies, distributed as
5,000 in region A, 14,000 in region B and 24,000 in region C.
After the cut the median, galaxy-weighted completeness is 88\%, 84\% and
88\% in regions A, B and C respectively.

We estimate $\xi(R,Z)$ using the \citet{LanSza93} estimator
\begin{equation}
  \xi(R,Z) = \frac{DD-2DR+RR}{RR}
\end{equation}
where $DD$, $DR$ and $RR$ are suitably normalized numbers of (weighted)
data-data, data-random and random-random pairs in bins of $(R,Z)$.
We experimented with two sets of weights, one to correct for fiber collisions
(described below) and one to reduce the variance of the estimator.  The
latter was 
\begin{equation}
  w_i = \frac{1}{1+\bar{n}(z_i)\bar{\xi}_V(s)}
\end{equation}
where $\bar{n}(z_i)$ is the mean density at redshift $z_i$ and $\bar{\xi}_V$
is a model for the volume-integrated redshift-space correlation function
within $s$.  We approximated $\bar{\xi}_V=4\pi s_0^2s$, corresponding to
$\xi(s)=(s/s_0)^{-2}$ and took $s_0=8\,h^{-1}$Mpc.
The details of the weighting scheme did not affect our final result on the
scales of interest to us here -- in fact dropping this weight altogether gave
comparable results and so we neglect this weight in what follows.

We are unable to obtain redshifts for approximately 7\% of the galaxies due to
fiber collisions -- no two fibers on any given observation can be placed closer
than $62''$.  At $z\simeq 0.5$ the $62''$ exclusion corresponds to
$0.4\,h^{-1}$Mpc.
Where possible we obtain redshifts for the collided galaxies in regions where
plates overlap, but the remaining exclusion must be account for.
We correct for the impact of this by (a) restricting our analysis to
relatively large scales and (b) up-weighting galaxy-galaxy pairs in the
analysis with angular separations smaller than $62''$.  The weight is
derived by comparing the angular correlation function of the entire
photometric sample with that of the galaxies for which we obtained redshifts
\citep{Haw03,Li06,Ros07}. This ratio is very close to unity above $62''$
but significantly depressed below this scale.
Note that in our situation there is a close correspondence between angular
separation and transverse separation since our survey volume is a relatively
narrow shell with reasonably large radius, so the number of pairs for which
this correction is appreciable is quite small.

Contours of the 2D correlation function for our $0.4<z<0.7$ galaxy sample
are shown in Figure \ref{fig:xiRZ}.  Note the characteristic elongation in
the $Z$ direction at small $R$ (fingers-of-god) and squashing at large $R$
(super-cluster infall).

To mitigate the effects of redshift space distortions we follow standard
practice and compute from $\xi(R,Z)$ the projected correlation function
\citep[e.g.][]{Dav83}
\begin{equation}
  w_p(R) = 2\int_0^\infty dZ\ \xi(R,Z) \qquad .
\label{eqn:wpdef}
\end{equation}
In practice we integrate to $100\,h^{-1}$Mpc, which is sufficiently large to
include almost all correlated pairs.
We also compute the angularly averaged, redshift space correlation function,
$\xi(s)$, and the cross-correlation between the CMASS sample selected from the
imaging and the spectroscopic samples, $w_\times$.
For all of these measures the full covariance matrix is computed from a set
of mock catalogs based on a halo-occupation distribution (HOD) modeling of the
data (\S\ref{sec:hod} and Appendix \ref{app:mocks}).

We now discuss each of the clustering measurements in turn, beginning
with the real-space clustering.

\subsection{Real-space clustering}

\begin{table*}
\begin{center}
\begin{tabular}{c|cccccccc}
     $R$ &   0.40  &   0.71  &   1.27  &   2.25  &   4.00  &   7.11 
 &  12.65  &  22.50 \\ \hline
 $R w_p$ & 167.68  & 134.49  & 147.63  & 168.29  & 208.77  & 242.70 
 & 255.89  & 230.36 \\ \hline
$\sigma$ &  15.91  &   6.26  &   6.54  &   7.87  &   9.89  &  14.16 
 &  20.49  &  28.77 \\ \hline
  0.40  &  1.000  &  0.266  &  0.185  &  0.216  &  0.202  &  0.174 
 &  0.139  &  0.168 \\
  0.71  &     --  &  1.000  &  0.346  &  0.329  &  0.312  &  0.299 
 &  0.238  &  0.202 \\
  1.27  &     --  &     --  &  1.000  &  0.580  &  0.533  &  0.561 
 &  0.487  &  0.371 \\
  2.25  &     --  &     --  &     --  &  1.000  &  0.695  &  0.652 
 &  0.552  &  0.417 \\
  4.00  &     --  &     --  &     --  &     --  &  1.000  &  0.793 
 &  0.703  &  0.522 \\
  7.11  &     --  &     --  &     --  &     --  &     --  &  1.000 
 &  0.826  &  0.646 \\
 12.65  &     --  &     --  &     --  &     --  &     --  &     -- 
 &  1.000  &  0.802 \\
 22.50  &     --  &     --  &     --  &     --  &     --  &     -- 
 &     --  &  1.000 
\end{tabular}
\end{center}
\caption{{\rm The projected correlation function data and covariance matrix,
for 8 equally spaced bins in $\ln R$.  Both $R$ and $w_p$ are measured
in units of $h^{-1}$Mpc, with $R$ quoted at the bin mid-point.
To reduce the condition number of the covariance matrix we quote means,
errors and covariances on $R w_p$, which removes much of the run of $w_p$
with scale and makes the quoted data points more similar in magnitude.
The error bars, $\sigma_i$, from the diagonal of $C$, are broken out separately
in the $3^{\rm rd}$ row and the correlation matrix, $C_{ij}/(\sigma_i\sigma_j)$
is quoted in the lower part of the table.  The full covariance matrix should
be used in any fit, and the finite width of the $R$ bins should be included in
theoretical predictions.}}
\label{tab:wp.004}
\end{table*}

\begin{figure}
\begin{center}
\resizebox{3.2in}{!}{\includegraphics{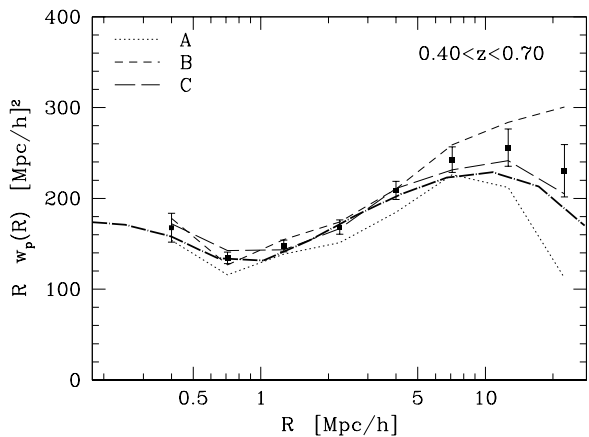}}
\end{center}
\caption{The projected correlation function for the $0.4<z<0.7$ sample
in regions A, B and C (lines) and for the combined sample (points with
errors).  The errors on the individual samples have been suppressed for
clarity.  The data are combined using the full covariance matrix, but only
the diagonal elements are plotted.
%The dot-dashed line shows a power-law correlation function with a correlation
The $w_p$ implied by a power-law correlation function of slope $-1.8$ and
correlation length of $7.5\,h^{-1}$Mpc forms a reasonable fit to the data with
$1<R_p<10\,h^{-1}$Mpc but we do not plot it here for clarity.
The (thick) long-dashed-dotted line shows the prediction of the best-fitting
HOD model (\S\protect\ref{sec:hod}), which provides a reasonable fit on all
scales plotted (recall the errors are correlated).}
\label{fig:wp}
\end{figure}

The projected correlation function for the $0.4<z<0.7$ sample is shown
in Figure \ref{fig:wp}.  We chose 8 bins, equally spaced in $\ln R$
between $0.3\,h^{-1}$Mpc and $30\,h^{-1}$Mpc as a compromise between
retaining the relevant information and generating stable covariance
matrices via Monte-Carlo.  The finite width of these bins should be
borne in mind when comparing theoretical models to these data.
The integration over $Z$ in Eq.~(\ref{eqn:wpdef}) was done by Riemann
sum using 100 linearly spaced bins in $Z$.  The results were well
converged at this spacing, because of the ``smearing'' of the correlation
function along the line-of-sight due to redshift space effects.
The data were analyzed separately in each of regions A, B and C and then
combined in a minimum variance manner:
\begin{equation}
  C^{-1} \mathbf{w}_p^{\rm (tot)} = \sum_{\alpha=A,B,C}
    \left[C^{(\alpha)}\right]^{-1} \mathbf{w}_p^{(\alpha)}
\end{equation}
with
\begin{equation}
  C^{-1} = \sum_\alpha \left[C^{(\alpha)}\right]^{-1}
\end{equation}
where $\mathbf{w}_p^{(\alpha)}$ represents the vector of $w_p$ measurements
{}from region $\alpha=$A, B or C.  Not surprisingly, the combined result is
dominated by the results from region C.
To reduce the condition number of the covariance matrix, and the dynamic range
in $w_p$, we fit throughout to $R\,w_p$ and quote the results in that form.
The $w_p$ points are quite covariant, in part because the integration in
Eq.~(\ref{eqn:wpdef}) introduces a large mixing of power at different $R$,
thus use of the full covariance matrix is essential.
The error bars on the individual $\mathbf{w}_p^{(\alpha)}$ have been
suppressed in the figure for clarity, and the square-root of the diagonal
elements of the covariance matrix are shown as error bars on the combined
result.

\begin{figure}
\begin{center}
\resizebox{3.2in}{!}{\includegraphics{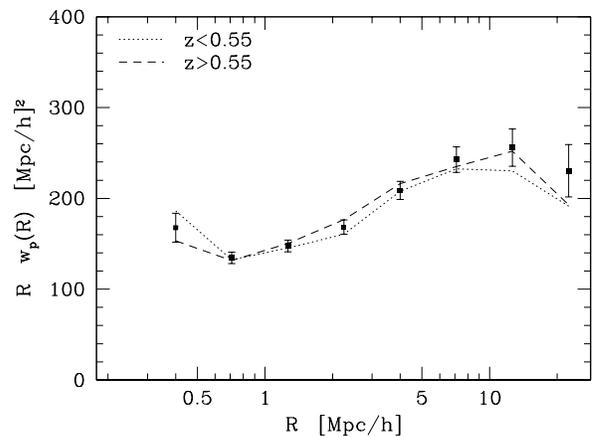}}
\end{center}
\caption{The projected correlation function of the high- and low-$z$ samples
(lines), split at the mid-point of the range, and of the full sample
(points with errors), indicating that the clustering is evolving little and
the sample can be analyzed in one wide redshift bin.}
\label{fig:wpz}
\end{figure}

We also subdivided the redshift range into a low-$z$ and high-$z$ half,
splitting at $z=0.55$, and found no statistically significant difference
between the two samples (Figure \ref{fig:wpz}; in the split samples the
fiber collision correction is more uncertain, so the disagreement at the
smallest $R$ point is not very significant).
This result motivates our decision to analyze the data in a single redshift
slice.
Slow evolution of the clustering is expected for a highly biased population
such as our luminous galaxies where the evolution of the bias approximately
cancels the evolution of the dark matter clustering \citep{Fry96}.

Even with only the 8 data points in $w_p$, deviations from a pure power-law
correlation function are apparent.  These can be traced to the non-power-law
nature of the mass correlation function and the way in which the galaxies
occupy dark matter halos -- we will return to these issues in \S\ref{sec:hod}.

The calculation of errors in clustering measurements can be done in a
number of different ways \citep[see][for discussion]{Nor09}.  We first
tried a bootstrap estimate, dividing the survey regions into 8-22,
roughly equal area ``pixels'' and sampling from these regions with replacement
\citep{Efron}.  Unfortunately the irregular geometry and relatively small
sky coverage meant we were not able to obtain a covariance matrix which was
stable against changes in the pixelization.  We anticipate that as the survey
progresses this technique will become more robust.  In the meantime, we
computed the covariance from a series of mock catalogs derived from an
iterative procedure using N-body simulations as described in
Appendix \ref{app:mocks}.
We will show in Figure \ref{fig:chi2} that the distribution
of $\chi^2$ from our mock catalogs encompasses the value obtained for the data
in regions A, B and C if both are computed using the mock-based covariance
matrix and the best-fitting HOD model (\S\ref{sec:hod}).
This indicates that the measurements we obtain are completely consistent with
being drawn from the underlying HOD model, given the finite number of galaxies
and observing geometry.

\subsection{Redshift-space clustering}

\begin{figure}
\begin{center}
\resizebox{3.2in}{!}{\includegraphics{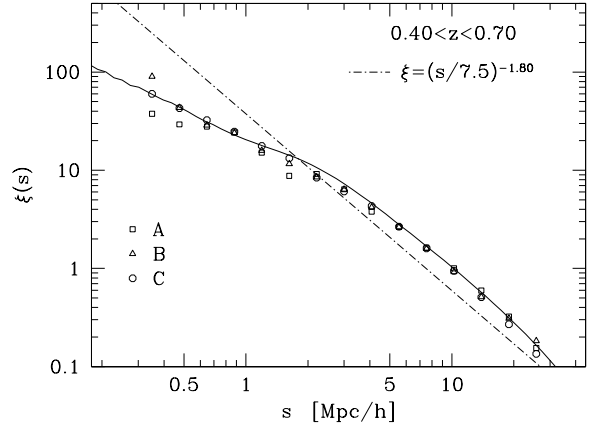}}
\end{center}
\caption{The redshift-space, isotropic correlation function for the
$0.4<z<0.7$ sample in regions A, B and C (points).  The same power-law
correlation function which fits the $w_p$ data on intermediate scales, with
$s_0=7.5\,h^{-1}$Mpc, is shown as the dot-dashed line while the solid
line is the prediction for $\xi(s)$ from the best-fitting HOD model
to $w_p$, assuming no velocity bias for satellites and that central galaxies
are at rest in their halos.
The good agreement below a few Mpc is an indication that the satellite fraction
in the model is close to that in the data and any velocity bias is small.}
\label{fig:xis}
\end{figure}

The angle-averaged redshift space correlation function, $\xi(s)$, for
the $0.4<z<0.7$ sample is shown in Figure \ref{fig:xis}.
Again, the data were analyzed separately in each of regions A, B and C.
The dot-dashed line shows the same power-law correlation function as described
in Figure \ref{fig:wp}, while the solid line shows the predicted $\xi(s)$
{}from the model that best fits the $w_p$ data (above).
The enhancement of clustering over the real-space result on large scales
(\citealt{Kai87}, for a review see \citealt{HamiltonReview} and for recent
developments see \citealt{PapSza08,ShaLew08})
is evident in the comparison of the data to the power-law.
The good agreement between the data and the HOD-model below a few Mpc is
indication that the satellite fraction in the model is close to that in
the data and the relative motions of the satellite galaxies are close to
the motions of the dark matter within the parent halos (i.e.~any velocity
bias is small).
The characteristic down-turn on scales smaller than a few Mpc is expected
{}from virial motions within halos and the motion of halos themselves.
The excess power of the HOD model compared to the data on scales of a few
Mpc can be mitigated by increasing the degrees of freedom in the model, for
example by dropping the assumption that central galaxies move with the mean
halo velocity or follow the dark matter radial profile or allowing a modest
amount of satellite velocity bias.

On scales below tens of Mpc the violations of the distant observer
approximation are small, but on larger scales they begin to become
appreciable \citep{PapSza08} and should be included in any comparison
between these data and a theoretical model (most noticeably for the
higher multipoles).
%A calculation and modeling of the higher-moments of $\xi(s)$, and the
%constraints on growth of structure, will be presented in a future
%publication.

\subsection{Cross-correlation}

Finally we consider $w_p$ computed from the cross-correlation
of the imaging catalog with the spectroscopy -- this allows us to
isolate the galaxies to a narrow redshift shell and convert
angles to (transverse) distances while at the same time being
insensitive to the details of the spectroscopic selection including
the issue of fiber collisions\footnote{One must still upweight some
of the spectroscopic galaxies to account for the fact that fiber collisions
occur more often in dense regions.  This issue turns out to be a very small
effect here, in part because BOSS is a deep survey and the correlation
between 2D over-density on the sky and 3D over-density is washed out by
projection.}.
As described in \citet{Pad09}, the angular cross correlation of the imaging
and spectroscopic samples, with angles converted to distances using the
redshift of the spectroscopic member, can be written as
\begin{equation}
  w_\times(R) = \left\langle f(\chi)\right\rangle\ w_p(R)
\label{eqn:w_cross}
\end{equation}
where $f(\chi)$ is the normalized radial distribution of the photometric
sample as a function of comoving distance, $\chi$, and the average is over
the redshift distribution of the spectroscopic sample.
Note that $w_\times(R)$ is dimensionless, with $f(\chi)$ having dimensions
of inverse length and $w_p$ having dimensions of length.

\begin{figure}
\begin{center}
\resizebox{3.2in}{!}{\includegraphics{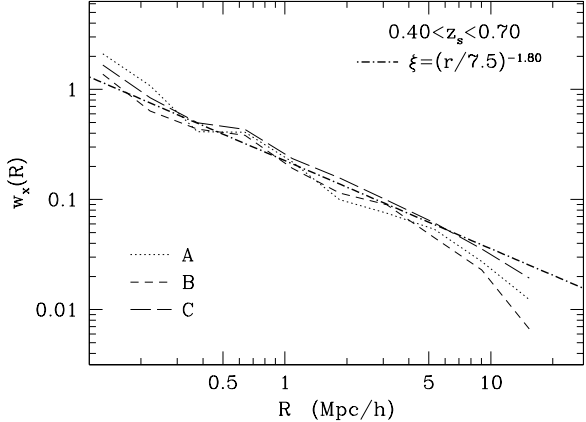}}
\end{center}
\caption{The cross-correlation function, $w_\times(R)$, of the spectroscopic
and photometric samples which is proportional to $w_p(R)$
(Eq.~\protect\ref{eqn:w_cross}).
The dot-dashed line represents a power-law correlation function.
Error bars have been suppressed to avoid obscuring the figure.
Due to the small value of $\langle f(\chi)\rangle\sim\mathcal{O}(10^{-3})$
the error bars are significant, especially at large scales, and are roughly
the difference between the plotted lines for regions A, B and C.}
\label{fig:wx}
\end{figure}

Figure \ref{fig:wx} shows the cross-correlation for regions A, B and C
along with a power-law correlation function.  The normalization of
this figure differs from that of Figure \ref{fig:wp} by a factor of
$\langle f(\chi)\rangle\sim\mathcal{O}(10^{-3})$.  Because the signal is
suppressed by the width of $f(\chi)$ the estimate of $w_p$ from the
cross-correlation is significantly noisier than that from the auto-correlation
\citep[see][\S2.1, for related discussion]{Mye09}.
The cross-correlation estimate is consistent with our auto-correlation results
but we have not attempted to fit any models to it directly.
We have extended the cross-correlation to smaller scales in the Figure to
emphasize that there is significant power even on very small scales, which
are difficult to probe directly with the auto-correlation function due to the
fiber collision problem.

\section{Halo occupation modeling} \label{sec:hod}

In order to relate the observed clustering of galaxies with the clustering
of the underlying mass, and to make realistic mock catalogs, we interpret
our measurements within the context of the halo occupation distribution
\citep{PeaSmi00,Sel00,Ben00,WHS01,BerWei02,CooShe02}.
The halo occupation distribution describes the number and distribution of
galaxies within dark matter halos.  Since the clustering and space density
of the latter are predictable functions of redshift, any HOD model makes
predictions for a wide range of observational statistics.
Rather than perform a simultaneous fit to the real- and redshift-space
correlation functions (including their covariances) we choose to fit to the
the real-space clustering only and show that the models which best fit these
data also provide a reasonable description of the redshift-space clustering
results.
This avoids the need to make additional assumptions for modeling the
redshift space correlation function.
We also implicitly assume that we are measuring a uniform sample of galaxies
across the entire redshift range, so that a single HOD makes sense.
We tested this assumption by splitting the sample into high- and low-redshift
subsamples.

We use a halo model which distinguishes between central and satellite
galaxies with the mean occupancy of halos:
\begin{equation}
  N(M)\equiv\left\langle N_{\rm gal}(M_{\rm halo})\right\rangle
  = N_{\rm cen}(M)+N_{\rm sat}(M)
  \quad .
\end{equation}
Each halo either hosts a central galaxy or does not, while the number of
satellites is Poisson distributed with a mean $N_{\rm sat}$.
The mean number of central galaxies per halo is modeled with\footnote{Note
that our definition of $\sigma$ can be interpreted as a fractional ``scatter''
in mass at threshold but is a factor $\ln(10)/\sqrt{2}$ different than that in
\protect\citet{Zhe05}.}
\begin{equation}
  N_{\rm cen}(M) = \frac{1}{2}
  \ {\rm erfc}\left[\frac{\ln(M_{\rm cut}/M)}{\sqrt{2}\sigma}\right]
\label{eqn:ncen}
\end{equation}
and
\begin{equation}
  N_{\rm sat}(M) = N_{\rm cen}
    \left(\frac{M-\kappa M_{\rm cut}}{M_1}\right)^\alpha
\label{eqn:nsat}
\end{equation}
for $M>\kappa M_{\rm cut}$ and zero otherwise.  This form implicitly assumes
that halos do not host satellite galaxies without hosting centrals, which
is at best an approximation, but this is reasonable for the purposes of
computing projected clustering.
Different functional forms have been proposed in the literature, but the
current form is flexible enough for our purposes.

\begin{figure}
\begin{center}
\resizebox{3.2in}{!}{\includegraphics{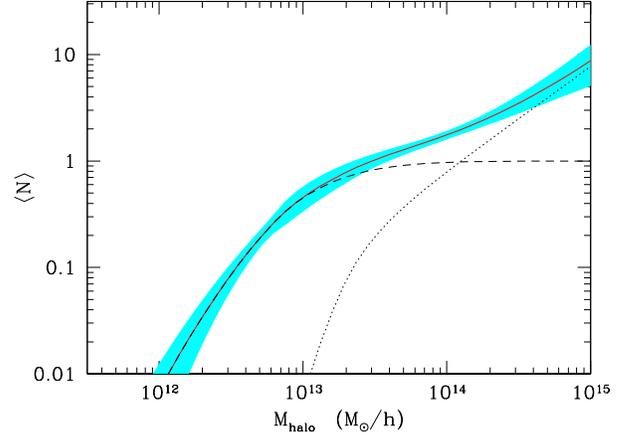}}
\end{center}
\caption{The mean occupancy of halos as a function of halo mass for our
full sample.  The shaded band indicates the $\pm 1\,\sigma$ range
determined from our Markov chain analysis.
The dashed and dotted lines show the average $N_{\rm cen}$ and $N_{\rm sat}$,
respectively.}
\label{fig:hod}
\end{figure}

To explore the plausible range of HOD parameter space we applied the
Markov Chain Monte Carlo method \citep[MCMC; e.g., see][]{Gil96} to the
$w_p$ data using a $\chi^2$-based likelihood.
This method generates a ``chain'' of HOD parameters whose frequency of
appearance traces the likelihood of that model fitting the data.
It works by generating random HODs from a trial distribution (in our case
a multi-dimensional Gaussian), populating a simulation cube with galaxies
according to that HOD, computing $w_p$ from the periodic box by pair counts
and accepting or rejecting the HOD based on the relative likelihood of the fit.
The step sizes and directions are determined from the covariance
matrix of a previous run of the chain.  Given the chain, the probability
distribution of any statistic derivable from the parameters can easily be
computed:
we show the mean occupancy of halos as a function of mass, $N(M)$, in
Figure \ref{fig:hod}, where the band indicates the $\pm 1\,\sigma$ spread
within the chain.  The mean (galaxy-weighted) halo mass is
$\langle M_{180b}\rangle_{\rm gal}=(2.8\pm 0.15)\times 10^{13}\,h^{-1}M_\odot$
(we quote here the mass interior to a sphere within which the mean density
is $180\times$ the background density for halo mass, rather than the
friends-of-friends mass, to facilitate comparison with other work);
while the satellite fraction is $(10\pm 2)\%$.
The values of the HOD parameters are given in Table \ref{tab:hod.004}.

In addition to the purely statistical errors, shown in the figure and quoted
above, there are systematic uncertainties.
Our correction for fiber collisions only significantly impacts the smallest
$R$ point in our calculation.  If we increase the error on that point by a
factor of 10, effectively removing it from the fit, the results change to
$\langle M_{180b}\rangle_{\rm gal}=(2.6\pm 0.15)\times 10^{13}\,h^{-1}M_\odot$
and $(7\pm 2)\%$ respectively which are shifts of approximately $1\,\sigma$.
Additional uncertainty arises from the uncertainty in the background cosmology
(held fixed in this paper) and from methodological choices.
A comparison of different methods for performing the halo modeling (using
different mass definitions or halo profiles, analytic vs.~numerical methods,
different ways of enforcing halo exclusion, etc.) suggests an additional
$\mathcal{O}(10\%)$ ``systematic'' uncertainty.  It would be interesting to
check the assumptions going into this HOD analysis, and the inferences so
derived, with additional data and a luminosity dependent modeling.

\begin{figure}
\begin{center}
\resizebox{3.2in}{!}{\includegraphics{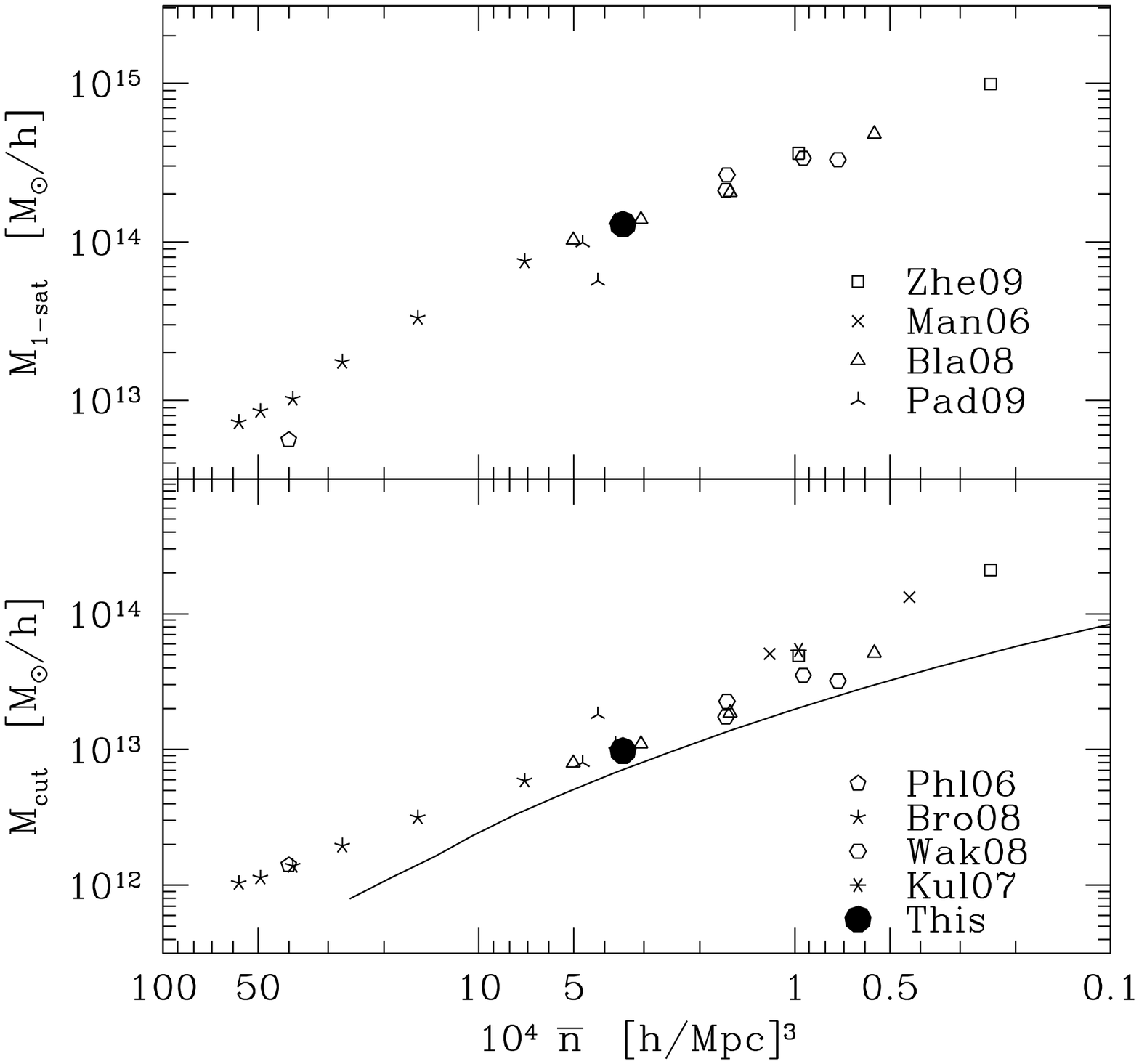}}
\end{center}
\caption{The HOD parameters, $M_{\rm cut}$ and $M_{1-{\rm sat}}$, as a
function of number density for a variety of intermediate redshift, massive
galaxy samples from the literature
(c.f.~Fig.~12 of \protect\citealt{Bro08}).
Here $M_{1-{\rm sat}}$ is the halo mass which hosts, on average,
one satellite which is easier to compare when different functional forms
for $N(M)$ are in use.
The data are taken from \protect\citet{Phl06}, \protect\citet{Man06},
\protect\citet{Kul07}, \protect\citet{Bla08}, \protect\citet{Bro08},
\protect\citet{Pad09}, \protect\citet{Wak08}, \protect\citet{Zhe09} and
this work, as noted in the legend.  Error bars on the individual points
have been suppressed for clarity, but are typically 0.1 dex.  The solid
line in the lower panel shows the halo mass function at $z=0.55$ for
comparison.  The value of $M_{\rm cut}$ for a sample of only central galaxies
with no scatter between observable and halo mass would follow this line.}
\label{fig:hod_compare}
\end{figure}

The halo occupancy of massive galaxies at these redshifts has been
investigated before based on both photometric
\citep{Whi07,Bla08,Bro08,Pad09}
and spectroscopic
\citep{Ros07,Ros08,Wak08,Zhe09,Rei09}
samples.
Accounting for differences in sample selection and redshift range, our
results appear quite consistent with the previous literature
(see Fig.~\ref{fig:hod_compare}).

\begin{table}
\begin{center}
\begin{tabular}{ccc}
 lg$M_{\rm cut}$ & $ 13.08 \pm   0.12$ & ($ 13.04$) \\
         lg$M_1$ & $ 14.06 \pm   0.10$ & ($ 14.05$) \\
        $\sigma$ & $  0.98 \pm   0.24$ & ($  0.94$) \\
        $\kappa$ & $  1.13 \pm   0.38$ & ($  0.93$) \\
        $\alpha$ & $  0.90 \pm   0.19$ & ($  0.97$) 
\end{tabular}
\end{center}
\caption{{\rm The mean and standard deviation of the HOD parameters
(see Eqs.~\protect\ref{eqn:ncen} and \protect\ref{eqn:nsat}) from our
Markov chain.  The particular values for our best-fit model are given
in parentheses.}}
\label{tab:hod.004}
\end{table}

Our galaxies populate a broad range of halo masses, with an approximate
power-law dependence of the mean number of galaxies per halo with
halo mass for massive halos and a broad roll-off at lower halo masses.
The low mass behavior is driven by the amplitude of the large-scale
clustering in combination with the relatively high number density of our
sample and encodes information about the scaling of the central galaxy
luminosity with halo mass and its distribution.
We find that the halos with masses $(2-3)\times 10^{13}\,h^{-1}M_\odot$
contain on average one of our massive galaxies.
At these redshifts such halos are quite highly biased (see below),
corresponding to galaxy groups, and we expect $b(z)\propto 1/D(z)$,
where $D(z)$ is the linear growth rate, leading to an approximately
constant clustering amplitude with redshift.

The majority of our galaxies are central galaxies residing in
$10^{13}\,h^{-1}M_\odot$ halos, but a non-negligible fraction are satellites
which live primarily in halos $\sim 10$ times more massive.  The width of
this ``plateau'' ($M_1/M_{\rm cut}$) is smaller than that found for less
luminous systems at lower redshift, though it continues the trends seen in
\citet{Zhe09} for plateau width and satellite fraction as a function of
luminosity.
This increase in the satellite fraction is driving the visibility of the
fingers-of-god in the correlation function
(Figure \ref{fig:xiRZ}) and the small-scale upturn in $w_p$.

\begin{figure}
\begin{center}
\resizebox{3.2in}{!}{\includegraphics{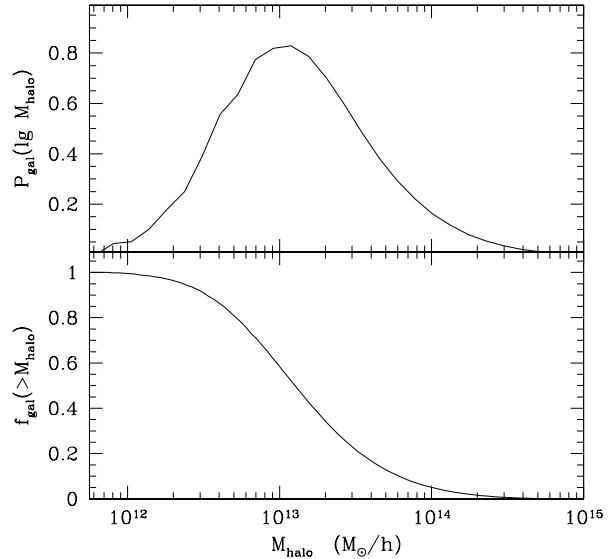}}
\end{center}
\caption{The probability per $\log_{10}M_{\rm halo}$ (upper) or cumulative
probability (lower) that a galaxy in our sample is hosted by a halo of mass
$M_{\rm halo}$.
Note the broad range of halo masses probed by our galaxies, and the low
probability of finding one of our galaxies in very high mass halos
(due to the sparsity of such halos at this redshift).}
\label{fig:hodcum}
\end{figure}

An alternative view of the halo occupation is presented in
Figure \ref{fig:hodcum}, which shows the probability that a
galaxy in our sample is hosted by a halo of mass $M$.
Note the broad range of halo masses probed by our galaxies, and the low
probability of finding one of our galaxies in very high mass halos --
which is a consequence of the sparsity of such halos at this redshift.

\begin{figure}
\begin{center}
\resizebox{3.2in}{!}{\includegraphics{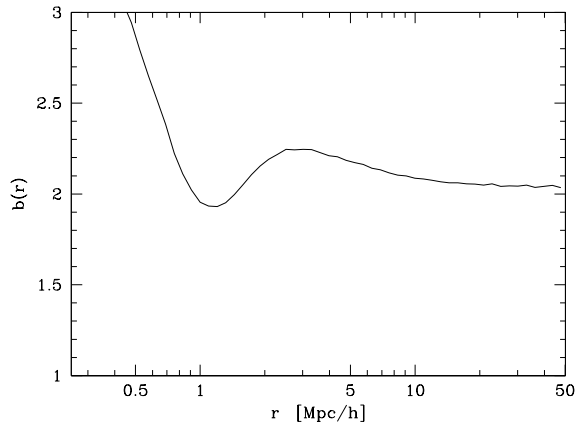}}
\end{center}
\caption{The scale-dependence of the bias,
$b(r)\equiv [\xi_{\rm gal}(r)/\xi_{\rm dm}(r)]^{1/2}$,
predicted from our best-fit halo model and N-body simulations.
The feature at a few Mpc has been seen in other analyses and occurs at
the transition between the 1- and 2-halo contributions (see text).
Note that the bias asymptotes to a constant, $b\simeq 2$, on large scales.}
\label{fig:bofr}
\end{figure}

The N-body simulations can also be used to infer the scale-dependence of the
bias, $b(r)\equiv [\xi_{\rm gal}(r)/\xi_{\rm dm}(r)]^{1/2}$, for our
best-fitting halo model.  This is shown in Figure \ref{fig:bofr},
where we see that above $10-20\,h^{-1}$Mpc the bias approaches a constant,
$b\simeq 2$.  For our cosmology the linear growth factor at $z=0.55$ is $0.762$
so $b\sigma_8(z=0.55)=1.3$ and this is assumed constant across our redshift
range.
This is very similar to the results obtained for photometric LRG samples at
comparable redshifts
\citep{Bla07,Ros07,Pad07,Pad09,Bla08}.
The rapid rise of $b(r)$ at very small scales is expected, since it is well
known that these galaxies exhibit an almost power-law correlation function
at small scales while the non-linear $\xi_{\rm dm}(r)$ is predicted to fall
below a power-law at small $r$.  (Most galaxy pairs on these scales are
central-satellite pairs, whereas for the dark matter there is no such
distinction so $\xi_{\rm dm}$ is the convolution of the halo radial profile
with itself.)
The feature in $b(r)$ at a few Mpc occurs at the transition between the
1- and 2-halo contributions, i.e.~pairs of galaxies that lie within
a single dark matter halo vs.~those which lie in separate halos, while
the rise at slightly larger scales comes from the scale-dependence of the
halo bias.
Note that the combination of the high clustering amplitude and number
density makes this sample particularly powerful for probing large-scale
structure at $z\simeq 0.5$.

Finally, using the best-fitting HOD model from the chain and a series of N-body
simulations we generate mock catalogs as described in more detail in
Appendix \ref{app:mocks}.
These are passed through the observational masks and cuts in order to mimic
the observations and can be analyzed in the same manner to generate a set
of mock measurements from which we compute covariance matrices and other
statistical quantities.  We match the redshift distribution of the sample to
our constant-number-density simulation boxes by randomly subsampling the
galaxies as a function of redshift
(with a 100\% sampling at the peak near $z\simeq 0.55$).
This is consistent with our assumption, earlier, that the HOD describes a
single population of objects and the $dN/dz$ reflects observational selection
effects.  We obtain similar HODs fitting to the thinner redshift slices which
lends credence to this view.

\section{Discussion} \label{sec:discussion}

The Baryon Spectroscopic Oscillation Survey is in the process of taking
spectra for 1.5 million luminous galaxies and 150,000 quasars to make
a precision determination of the scale of baryon oscillations and to
study the growth of structure and the evolution of massive galaxies.
We have presented measurements of the clustering of 44,000 massive galaxies
at $z\sim 0.5$ from the first semester of BOSS data,
showing that BOSS is performing well and that the galaxies we are targeting
have properties in line with expectations \citep{SchWhiEis09}.

The CMASS sample at $z\simeq 0.5$ has a large-scale bias of $b\simeq 2$
(Fig.~\ref{fig:bofr}), and a number density several times higher than
the earlier, spectroscopic LRG sample of \citet{Eis01}, making it an ideal
sample for studying large-scale structure.
The majority of our CMASS galaxies are central galaxies residing in
$10^{13}\,h^{-1}M_\odot$ halos, but a non-negligible fraction are satellites
which live primarily in halos $\sim 10$ times more massive.

The data through July 2010 do not cover enough volume to robustly detect the
acoustic peak in the correlation function in this sample, one of the science
goals of BOSS.
%Fig.~\ref{fig:xis_now} shows the mean and scatter in $\xi(s)$ from our mock
%catalogs.
While no definitive detection is possible at present, the error bars are
anticipated to shrink rapidly as we collect more redshifts; BOSS should
be able to constrain the acoustic scale at $z\sim 0.5$ within the next year,
with the constraints becoming increasingly tight as the survey progresses.

\vspace*{1cm}

\begin{acknowledgments}
  Funding for SDSS-III has been provided by the Alfred P. Sloan Foundation,
  the Participating Institutions, the National Science Foundation, and the
  U.S. Department of Energy. The SDSS-III web site is http://www.sdss3.org/.

  SDSS-III is managed by the Astrophysical Research Consortium for the
  Participating Institutions of the SDSS-III Collaboration including the
  University of Arizona, the Brazilian Participation Group, Brookhaven
  National Laboratory, University of Cambridge, University of Florida,
  the French Participation Group, the German Participation Group, the
  Instituto de Astrofisica de Canarias, the Michigan State/Notre Dame/JINA
  Participation Group, Johns Hopkins University, Lawrence Berkeley National
  Laboratory, Max Planck Institute for Astrophysics, New Mexico State
  University, New York University, the Ohio State University, the Penn
  State University, University of Portsmouth, Princeton University,
  University of Tokyo, the University of Utah, Vanderbilt University,
  University of Virginia, University of Washington, and Yale University.

  The analysis made use of the computing resources of the
  National Energy Research Scientific Computing Center,
  the Shared Research Computing Services Pilot of the
  University of California and
  the Laboratory Research Computing project at
  Lawrence Berkeley National Laboratory.

  M.W. was supported by the NSF and NASA.
  F.P. acknowledges support from the Spanish MICINN's Consolider
  grant MultiDark CSD2009-00064.
\end{acknowledgments}

\appendix

\section{N-body simulations and mock catalogs} \label{app:mocks}

We make use of several simulations in this paper.  The main set is 20
different realizations of the $\Lambda$CDM family with $\Omega_m=0.274$,
$\Omega_\Lambda=0.726$, $h=0.7$, $n=0.95$ and $\sigma_8=0.8$ (in agreement
with a wide array of observations).
Briefly, each simulation employs an updated version of the TreePM code
described in \citet{TreePM} to evolve $1500^3$ equal mass
($7.6\times 10^{10}\,h^{-1}M_\odot$)
particles in a periodic cube of side length $1500\,h^{-1}$Mpc with a
Plummer equivalent smoothing of $36\,h^{-1}$kpc.
The initial conditions were generated by displacing particles from a regular
grid using second order Lagrangian perturbation theory at $z=75$ where the
rms displacement is $10\%$ of the mean inter-particle spacing.
This TreePM code has been compared to a number of other codes and shown to
perform well for such simulations \citep{Hei08}.
Recently the code has been modified to use a hybrid MPI+OpenMP approach which
is particularly efficient for modern clusters.

For each output we found dark matter halos using the Friends of Friends (FoF)
algorithm \citep{DEFW} with a linking length of $0.168$ times the mean
interparticle spacing.  This partitions the particles into equivalence
classes roughly bounded by isodensity contours of $100\times$ the mean
density.  The position of the most-bound particle, the center of mass velocity
and a random subset of the member particles are stored for each halo and
used as input into the halo occupation distribution modeling and mock
catalogs.  Throughout we use the sum of the masses of the particles linked by
the FoF algorithm as our basic definition of halo mass, except when
quoting $\langle M_{180b}\rangle_{\rm gal}$ in \S\ref{sec:hod} where we use
spherical over-density (SO) masses to facilitate comparison with other work.
Note we do not run a SO finder to define new groups.  We use the FoF halo
catalog, only computing a different mass for each FoF halo.
In order to compute these SO masses we grow spheres outwards from the most
bound particle in each FoF halo, stopping when the mean density of the
enclosed material (including both halo and non-halo particles) is $180\times$
the background density.
The total enclosed mass we denote by $M_{180b}$.

\begin{figure*}
\begin{center}
\resizebox{5.5in}{!}{\includegraphics{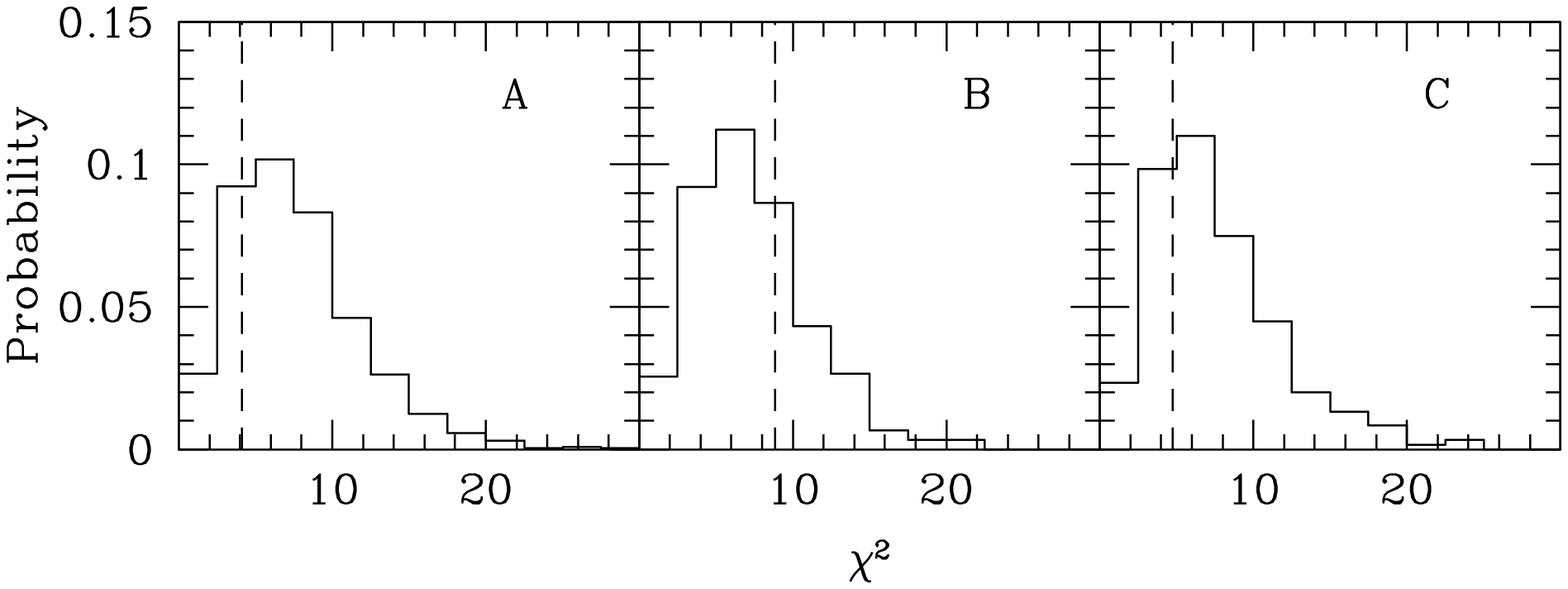}}
\end{center}
\caption{The distribution of $\chi^2$ for $w_p$ from our mock catalogs
(histograms) and from the data (vertical dashed lines) in regions A, B and C.
The $\chi^2$ is computed for the measured or mock $w_p$ compared to the
best-fitting HOD model using the covariance matrix computed from the mocks.
The measurements we obtain are completely consistent with being drawn from
the underlying HOD model, given the finite number of galaxies and observing
geometry.}
\label{fig:chi2}
\end{figure*}

All of the mock observational samples are assumed to be iso-redshift, and
``static'' outputs are used as input to the modeling.  The assumption of
non-evolving clustering over the relevant redshift range is theoretically
expected for a highly biased population, and also borne out by our
modeling (\S\ref{sec:hod}) and measurements (\S\ref{sec:clustering}).

Once a set of HOD parameter values has been chosen, we populate each halo in
a given simulation with mock ``galaxies''.
The HOD provides the probabilities that a halo will contain a central galaxy
and the number of satellites.  The central galaxy is placed at the position
of the most bound particle in the halo, and we randomly draw dark matter
particles to represent the satellites, assuming that the satellite galaxies
trace the mass profile within halos.
This approach has the advantage of retaining any alignments between the halo
material, the filamentary large-scale structure and the velocity field.

Since the observational geometry is in some cases highly elongated
(Fig.~\ref{fig:sky}), we use volume remapping \citep{Car10} on the periodic
cubes to encompass many realizations of the sample within each box.
The mock galaxies are then observed in a way analogous to the actual sample,
with the completeness mask and redshift cuts applied to generate several
hundred ``mock surveys''.
(Overall we have $1500$ mock surveys, divided into $900$, $360$ and $240$
mock surveys of regions A, B and C respectively.  However they are not all
completely independent as we have only $\sim 100$ times as much volume in
the simulations as in the largest region, C.)
For technical reasons, and since it only affects the smallest scale $w_p$
point, we do not model fiber collisions.  Instead we increase the errors for
that point by the square root of the ratio of the pair counts in the
photometric sample to that in the spectroscopic sample
(i.e.~the same correction applied to the data-data pairs in computing
$\xi(R,Z)$).
This correction is appropriate in the limit that the error is dominated by
Poisson counting statistics.

The covariance matrices for the clustering statistics are obtained from the
mocks, and the entire procedure (reconstructing the best-fit with the
new covariance matrix, recomputing the mock catalogs and recomputing the
clustering) is iterated until convergence.  Given a reasonable starting HOD,
the procedure converges within two or three steps.

Over the range of scales probed in this paper the correlation function is
quite well constrained and we find the distribution of $w_p$ values in
the mocks is well fit by a Gaussian at each $R$.  This suggests we are
able to use a Gaussian form for the likelihood, which is backed up by the
distribution of $\chi^2$ values seen in Figure \ref{fig:chi2}.

\end{document}